\documentclass[journal]{IEEEtran}

\ifCLASSINFOpdf
  \usepackage[pdftex]{graphicx}
  \graphicspath{{./}{Figs/}}
  \DeclareGraphicsExtensions{.pdf,.jpeg,.png,.eps}
\else
\fi
\usepackage{amssymb}
\usepackage{amsmath}
\usepackage{mathtools}
\usepackage{bigdelim}

\begin{document}

\title{Stable Throughput Region of Cognitive-Relay Networks with Imperfect Sensing and Finite Relaying Buffer}

\author{Ahmed~M.~Alaa 
\thanks{The authors are with XXXX.}
\thanks{Manuscript received XXXX XX, 201X; revised XXXX XX, 201X.}}

\maketitle
\begin{abstract}
In this letter, we obtain the stable throughput region for a cognitive relaying scheme with a finite relaying buffer and imperfect sensing. The analysis investigates the effect of the secondary user's finite relaying capabilities under different scenarios of primary, secondary and relaying links outages. Furthermore, we demonstrate the effect of miss detection and false alarm probabilities on the achievable throughput for the primary and secondary users.
\end{abstract}

\section{Introduction}

The paradigm of cognitive radio promises a more efficient utilization of the scarce spectrum occupied by the non-permanently active primary (licensed) users [7]. In a conventional cognitive spectrum sharing scheme, a secondary user (SU) senses the spectrum to check the activity of a primary user (PU). If the PU is idle, the SU opportunistically utilizes its allocated frequency bands for its own data transmission. Moreover, the SU can offer cooperative relaying for the PU packets if the PU link suffers from a high outage probability [1]. In such scheme, the SU receives the packets that are unsuccessfully decoded by the primary destination [8], increasing the availability of the spectrum for its own data transmission and increasing the probability of successful PU transmission. 

In a practical system, the SU will sense the existence of the PU in an imperfect manner. This means that the SU operation would perturb the achievable rate of a PU with exclusive spectrum access. In [2], the stable throughput region was characterized for a shared channel where the SU randomly accesses the channel while relaying unsuccessfully transmitted PU data. However, the relaying queue was assumed to be an M/M/1 queue with infinite buffering capability. In [1], cooperative relaying was incorporated with the original notion for cognitive radio where the SU is allowed to access the channel only when the PU is idle. Again, the SU was assumed to buffer any amount of primary data. Similar analyses incorporated the physical layer parameters to determine the outage probabilities in [3] and [4]. A generalization for the cognitive scheme was studied in [5], where multiple secondary users exists, but all are equipped with unlimited relaying buffers. 

\begin{figure}[!t]
\centering
\includegraphics[width=2 in]{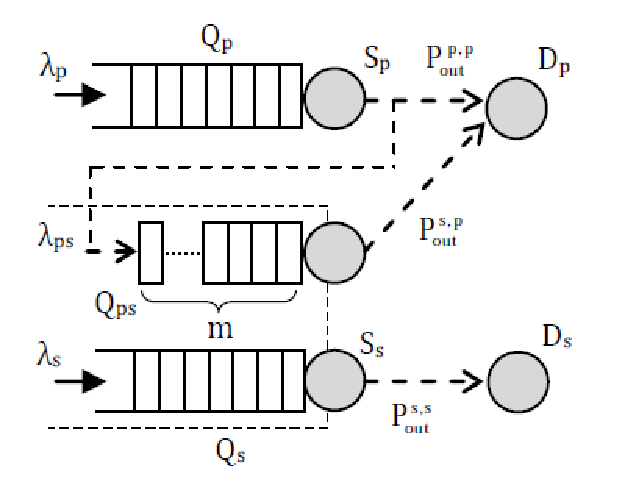}
\caption{Queue model for cognitive-relay scheme with finite relaying buffer.}
\label{fig_sim}
\end{figure}

In this work, we focus on the impact of the size of the relaying queue on the stable throughput region of cognitive-relay networks. The impact of relaying size is found to depend on the outage probabilities of primary and secondary links. Moreover, we characterize the impact of the false alarm and detection probabilities on the stable throughput region.          

The rest of the paper is organized as follows: the system model is presented in section II. In section III, we derive the stable-throughput region. Numerical results are presented in section IV. Finally, we draw our conclusion in section V.  

\section{System Model}

The system under study is a conventional cognitive system where the SU accesses the channel only when the PU is idle [1]. The queuing model of such system is demonstrated in Fig. 1. The primary transmitter $S_{p}$ is equipped with a primary queue $Q_{p}$ which buffers primary data arriving at a rate of $\lambda_{p}$ and departing at a rate of $\mu_{p}$. Similarly, the secondary transmitter $S_{s}$ is equipped with a secondary queue $Q_{s}$ with parameters $\lambda_{s}$ and $\mu_{s}$. Besides, the secondary users employs a seperate relaying queue $Q_{ps}$ that holds the relayed data arriving and leaving with rates of $\lambda_{ps}$ and $\mu_{ps}$ respectively. The relaying queue has a limited length of $m$.

The system is assumed to be time slotted, with the slot duration accomodating one packet. All the arrival and departure rates specified are in packets/slot. The primary and secondary destinations $D_{p}$ and $D_{s}$ are assumed to send immediate acknowledgments (ACK) on a perfect channel upon successful reception. While relaying the PU data, the SU emulates a primary destination and sends an ACK to the PU allowing it to drop the packet of interest from $Q_{p}$. The physical links $S_{p}$-$D_{p}$, $S_{p}$-$S_{s}$, $S_{s}$-$D_{p}$, and $S_{s}$-$D_{s}$ are characterized by the outage probabilities $P_{out}^{p,p}$, $P_{out}^{p,s}$, $P_{out}^{s,p}$ and $P_{out}^{s,s}$, respectively. 

The SU senses the PU activity (current queue length) with a probability of detection $P_{d}$ and a false alarm probability of $P_{f}$. We assume that simultaneous transmissions from both PU and SU leads to a collision that corrupts both packets. The PU accesses the channel whenever its queue is not empty, whereas the SU accesses the channel only when the primary user has no packets in its queue.

\section{Stable Throughput Region}

Our objective is to characterize the stable throughput region of the presented cognitive-relay scheme taking into consideration. Since the mean service rates at $S_{s}$ and $S_{p}$ depend on each other's queue size, these queues are called interacting, and consequently the rates of the individual departure processes cannot be computed directly. In order to bypass this problem, we utilize the idea of stochastic dominance, which has been employed before in [6] to analyze interacting queues. In this setting, the SU is assumed to be always backlogged. In other words, we construct a dominant system where the SU transmits dummy packets whenever its queue is empty. Thus, all collision events in the original system occurs in the dominant system. 

In the stochastic dominance approach, we first construct an appropriate dominant system, which is a modification of the original system, that ensures that the queue sizes in the dominant system are, at all times, at least as large as those of the original system. Thus, the stability region of the new system "inner bounds" that of the original system. Furthermore, in the new system the queues are decoupled and not interacting anymore, thereby permitting the characterization of the stability region.

We start by characterizing the service rate $\mu_{p}$ (probability of successful transmission) perceived by the primary queue

\begin{equation} 
\label{eqn_example} 
\mu_{p} = P_{d} ((1-P_{out}^{p,p}) + P_{out}^{p,p} (1-P_{out}^{p,s}).P(Q_{ps} \neq m) ),
\end{equation} 

which corresponds to the probability that SU detects the PU and the PU channel is either not in outage, or outaged but with a PU to SU relaying in a non-full buffer. By defining the loading factor for the M/M/1/m relaying queue as $\rho_{ps}$ = $\frac{\lambda_{ps}}{\mu_{ps}}$, the probability of no overflow for the relaying buffer is

\[ P(Q_{ps} \neq m) = 1 - \frac{1-\rho_{ps}}{1-\rho_{ps}^{m+1}} \rho_{ps}^m.\]

Thus, the primary user's service rate is given by

\begin{equation} 
\label{eqn_example} 
\mu_{p} = P_{d} ((1-P_{out}^{p,p}) + P_{out}^{p,p} (1-P_{out}^{p,s}) (1 - \frac{1-\rho_{ps}}{1-\rho_{ps}^{m+1}} \rho_{ps}^m).
\end{equation} 

The arrival rate of the relayed packets depend on the outage of the primary channel, the probability of relaying buffer overflow and the probability of an empty PU buffer. The relayed data arrival rate can be given by:

\[\lambda_{ps} = P(Q_{ps} \neq m).P(Q_{p} \neq 0).P_{d}.(1-P_{out}^{p,s}). P_{out}^{p,p}, \]

which reduces to

\begin{equation} 
\lambda_{ps} = (1 - \frac{1-\rho_{ps}}{1-\rho_{ps}^{m+1}} \rho_{ps}^m).\frac{\lambda_{p}}{\mu_{p}}.P_{d}.(1-P_{out}^{p,s}). P_{out}^{p,p}.
\end{equation} 

Note that to obtain the value of $\lambda_{ps}$, we have to solve a transcendental equation. This results from the usage of an M/M/1/m queue rather than an infinite relaying queue. The relaying queue service rate depends on the probability of an empty primary queue, and the outage and false alarm probabilities as

\begin{equation} 
\mu_{ps} = (1-\frac{\lambda_{p}}{\mu_{p}}).P_{d}.(1-P_{out}^{s,p}). (1-P_{f}).
\end{equation} 

Similarly, the secondary user's service rate can be formulated as

\begin{equation} 
\mu_{s} = (\frac{1-\rho_{ps}}{1-\rho_{ps}^{m+1}}).(1-\frac{\lambda_{p}}{\mu_{p}}).P_{d}.(1-P_{out}^{s,s}). (1-P_{f}).
\end{equation} 

The stable throughput region is obtained by applying the Loynes' theorem. For the case of the relaying queue, it is always stable because of its finite length. Based on Loynes' theorem, we have that $\lambda_{p} \textless \mu_{p}$ and $\lambda_{s} \textless \mu_{s}$. By defining the following auxiliary parameters $\psi = (1-P_{f}).(1-P_{out}^{s,p})$, $\eta = (1-P_{out}^{p,p}.P_{out}^{s,p})$, and $\phi = \frac{(1-P_{out}^{p,s}).P_{out}^{p,p}}{(1-P_{out}^{p,p}.P_{out}^{p,s})}.$ The stable throughput region is obtained by scanning $\rho_{ps}$ over the range [0, 1] and finding the resulting pair of ($\lambda_{p}$, $\lambda_{s}$) by solving the following transcendental equations:

\[ \lambda_{s} \textless \frac{1-\frac{\phi \lambda_{p}}{\psi (1-\frac{\lambda_{p}}{\mu_{p}})}}{1-(\frac{\phi \lambda_{p}}{\psi (1-\frac{\lambda_{p}}{\mu_{p}})})^{m+1}} .(1-\frac{\lambda_{p}}{\mu_{p}}) .(1-P_{f}).(1-P_{out}^{s,s}),\]

\[ \lambda_{p} \textless P_{d} ((1-P_{out}^{p,p}) + P_{out}^{p,p} (1-P_{out}^{p,s}) (1 - \frac{1-\rho_{ps}}{1-\rho_{ps}^{m+1}} \rho_{ps}^m)\]

and

\[ \rho_{ps} = \frac{\lambda_{p}}{\mu_{p} - \lambda_{p}} \frac{(1 - \frac{1-\rho_{ps}}{1-\rho_{ps}^{m+1}} \rho_{ps}^m).P_{d}.(1-P_{out}^{p,s}). P_{out}^{p,p}}{(1-P_{out}^{s,p}). (1-P_{f})} \]

\section{Numerical Results}

\begin{figure}[!t]
\centering
\includegraphics[width=2 in]{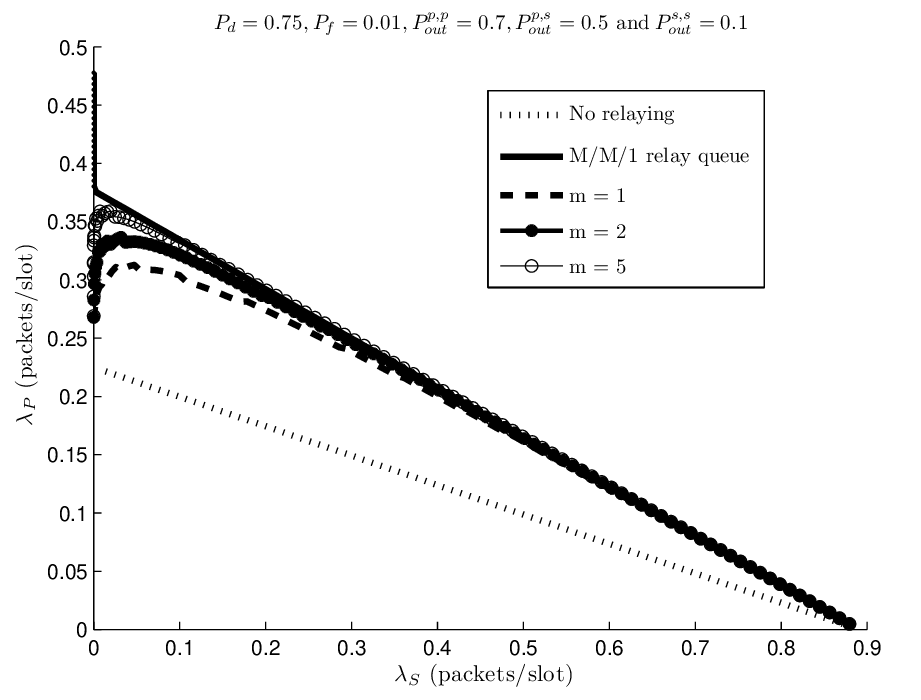}
\caption{Stable throughput region with high outage on the primary link.}
\label{fig_sim}
\end{figure}

\begin{figure}[!t]
\centering
\includegraphics[width=2 in]{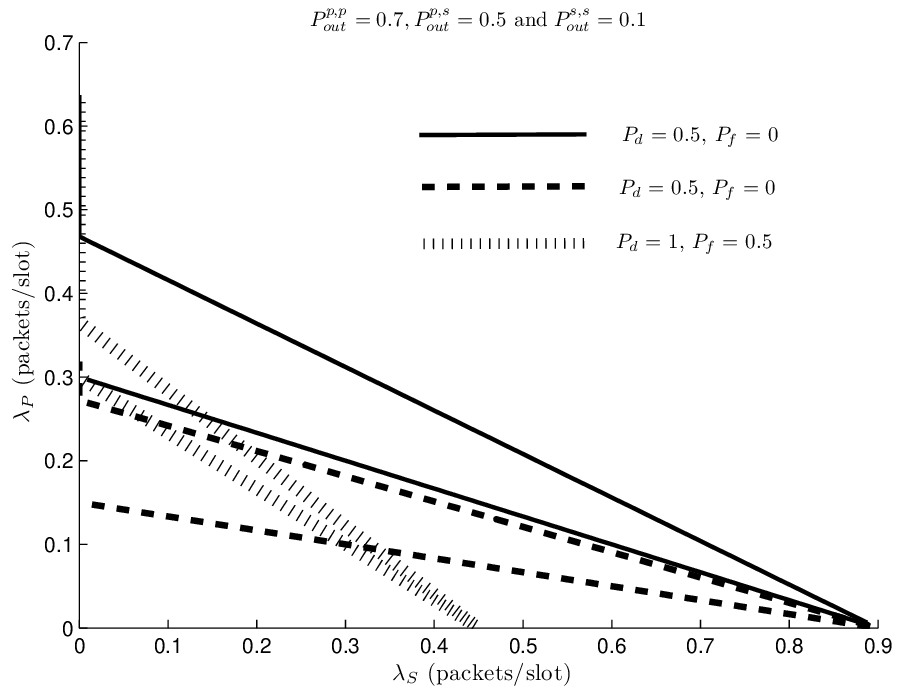}
\caption{Impact of imperfect sensing on the stable throughput region.}
\label{fig_sim}
\end{figure}

\section{Impact of Finite Relaying Buffer}

An interesting fact is that a very small relaying buffer is sufficient to achieve the same throughput of an M/M/1 relaying queue. In Fig. 2, we plot the achievable rates with m = 1, 2 and 5. It is shown that all these buffer lengths, despite of being relatively small, are sufficient to expand the PU stable throughput region in a similar manner to the M/M/1 queue. 

\section{Impact of Imperfect Sensing}

Fig. 3 depicts the impact of the imperfect sensing on the stable throughput boundaries for both cooperative and non-cooperative cases. We note that the imperfect probability of detection causes deterioration for the achievable primary user rate. This is due to the fact that miss detection causes sure collisions between primary ans secondary users. The deterioration is nearly proportional to the value of $P_{d}$. On the other hand, the false alarm probability causes loss of both primary and secondary data rates. An interesting fact is that in case of no-relaying the false alarm probability does not affect the achievable rate for the PU. This is because when no cooperation, a false alarm will not cause a transmission opportunity loss to the PU but will cause a loss to the SU. However, when relaying is applied, a false alarm will prevent the SU from forwarding the relayed data and neither the PU nor the SU utilize the channel causing a degradation in both rates. Thus, we recommend using a Neyman-pearson test for sensing, where the false alarm probability is set to a certain value dictated by the desired secondary rate and maximizing the achievable PU rate. 

\section{Conclusion}
In this letter, we obtained the stable throughput region for a cognitive relaying scheme with a finite relaying buffer and imperfect sensing. The analysis investigated the impact of the secondary user's finite relaying capabilities under different scenarios of primary, secondary and relaying links outages. Furthermore, we demonstrated the effect of miss detection and false alarm probabilities on the achievable throughput for the primary and secondary users. It was found that detection probability affects the PU achievable rates only, while false alarm probability can affect the SU rates if the SU offers relaying to the PU. Besides, it was shown that a relaying buffer that is as small as one packet sized can achieve almost the same rates of an infinite M/M/1 queuing buffer.

\end{document}